\documentclass[final,10pt]{elsart1p}

\usepackage{graphicx}

\usepackage{amssymb}

\begin{document}

\begin{frontmatter}



\title{Common aspects of q-deformed Lie algebras and fractional calculus}


\author{Richard Herrmann}
\ead{herrmann@gigahedron.com}

\address{GigaHedron, Berliner Ring 80, D-63303 Dreieich, Germany}

\begin{abstract}
Fractional calculus and q-deformed Lie algebras are closely related. 
Both concepts expand the scope of standard Lie algebras to describe generalized symmetries. 
A new class of fractional q-deformed Lie algebras is proposed, which for the first time allows a smooth transition between different Lie algebras.

For the fractional harmonic oscillator, the corresponding fractional q-number is derived. 
It is shown, that the resulting energy spectrum is an appropriate tool to describe e.g. the ground state spectra of even-even nuclei. 
In addition, the equivalence of rotational and vibrational spectra 
for fractional q-deformed Lie algebras is shown and the $B_\alpha(E2)$ values
for the fractional q-deformed symmetric rotor are calculated.

A first interpretation of half integer representations of the fractional rotation group is given in terms of a  description of $K=1/2^-$ band spectra of odd-even nuclei.

\end{abstract}

\begin{keyword}
Perturbation and fractional calculus methods \sep
q-deformed Lie-algebras

\PACS 21.60.Fw \sep 21.10.k
\end{keyword}
\end{frontmatter}

\section{Introduction}
The combination of concepts and methods, developed in different branches of physics,
has always led to new insights and improvements. In nuclear physics for example, the
description of rotational and vibrational nuclear spectra has undoubtedly been influenced by
concepts, which were first developed for molecular spectra.

The intention of this paper is similar. We will show, that the concept of q-deformed
Lie algebras and the methods developed in fractional calculus are closely related and 
may be combined leading to 
a new class of fractional q-deformed Lie algebras.  

An interesting  question of physical relevance is whether q-deformed  Lie algebras are not only suitable for describing small deviations
from Lie symmetries, but in addition can bridge different Lie symmetries \cite{bon}. We will demonstrate that fractional q-deformed
Lie algebras show exactly this behaviour.
For that purpose, we will first introduce the necessary tools for a fractional extension of the usual q-deformed
Lie algebras and present 
the fractional analogue of the standard q-deformed oscillator. 

As an application  we will use the fractional q-deformed oscillator for a description of   
the low energy excitation
ground state band spectra of even-even nuclei.
In addition, the equivalence of rotational and vibrational spectra 
for the fractional q-deformed Lie algebras is demonstrated and  the $B_\alpha(E2)$ values
for the fractional q-deformed symmetric rotor are discussed.

Finally we present a first application of half integer representations of the fractional rotation group $SO^\alpha(3)$ in terms of a
successful interpretation of the $K=1/2^-$  band spectrum of the nucleus $^{183}$W$_{74}$.

\section{q-deformed Lie algebras}
q-deformed Lie algebras are extended versions of the usual Lie algebras \cite{bon}. They provide us with an extended set 
of symmetries and therefore allow the description of physical phenomena, which are beyond the scope of usual 
Lie algebras.
In order to describe a q-deformed Lie algebra, we introduce a parameter $q$ and define a mapping of 
ordinary numbers $x$ to q-numbers  via:
\begin{equation}
\label{q0}
\left[ x \right]_q = {q^x - q^{-x} \over q - q^{-1}} 
\end{equation}
which in the limit ${q \rightarrow 1}$ yields the ordinary numbers $x$
\begin{equation}
\label{qlimit}
\lim_{q \rightarrow 1}\left[ x \right]_q = x 
\end{equation}
and furthermore we obtain as a special value
\begin{equation}
\label{q00}
\left[ 0 \right]_q = 0 
\end{equation}

Based on q-numbers a q-derivative may be defined via:
\begin{equation}
\label{qd}
D_x^q f(x)  = {f(q x) - f(q^{-1}x)\over (q-q^{-1})x} 
\end{equation}
With this definition for a function $f(x) =  x^n$ we get
\begin{equation}
\label{qdx}
D_x^q \, x^n  =  \left[n \right]_q x^{n-1} 
\end{equation}
As an  example for q-deformed Lie algebras we introduce the q-deformed harmonic oscillator. The
creation and annihilation operators $a^\dagger$, $a$ and the number operator $N$ generate the algebra:
\begin{eqnarray}
\left[ N, a^\dagger \right] &=& a^\dagger \\
\left[ N, a \right] &=& -a \\
\label{n3}
a a ^\dagger - q^{\pm 1} a^\dagger a &=& q^{\mp N}
\end{eqnarray} 
with  (\ref{q0}) an alternative definition of (\ref{n3}) is given by
\begin{eqnarray}
a^\dagger a &=& \left[N\right]_q \\
a a^\dagger &=& \left[ N+1 \right]_q 
\end{eqnarray}
Defining a vacuum state with $a |0> = 0$, the action of the operators $\{a,a^\dagger,N\}$ on the basis $|n>$ of a Fock space, 
defined by a repeated action of the creation operator on the vacuum state,
is given by:
\begin{eqnarray}
\label{n}
N |n> &=& n |n> \\
a^\dagger |n> &=& \sqrt{\left[ n+1 \right]_q} |n+1>\\
a |n> &=& \sqrt{\left[ n \right]_q} |n-1>
\end{eqnarray} 
The Hamiltonian of the q-deformed harmonic oscillator is defined as
\begin{equation}  
H = {\hbar \omega \over 2 }(a a^\dagger + a^\dagger a)
\end{equation}
and the eigenvalues on the basis $|n>$ result as
\begin{equation}
\label{eho}  
E^q(n) = {\hbar \omega \over 2 }(\left[ n \right]_q + \left[ n+1  \right]_q )
\end{equation}
\section{Fractional calculus}
During the last  years the interest of physicists in non-local field theories has been steadily increasing. The main reason for this development is the expectation, that the use of these field theories will lead to a much more elegant and effective way of treating problems in particle and high-energy physics, as it has been possible up to now with local field theories.

A particular subgroup of non-local field theories is described with operators of fractional nature and is specified within the framework of fractional calculus.
Fractional calculus  \cite{ol}-\cite{ff} provides us with a set of axioms and methods to extend the concept of a derivative operator from integer order $n$ to arbitrary order $\alpha$, 
where $\alpha$ is a real value: 
\begin{equation}
\{ x^n, {\partial^n \over \partial x^n} \} 
\rightarrow
\{ x^\alpha, {\partial^\alpha \over \partial x^\alpha} \}
\end{equation}
Despite the fact, that this concept has been discussed since the days of Leibniz and since then has occupied the great mathematicians of their times, no other research area has resisted a direct application for centuries. Consequently, Abel’s treatment of the tautochrone problem from 1823 \cite{abel}  stood for a long time as a singular example for an application of fractional calculus. 
Not until the works of Mandelbrot  on fractal geometry  \cite{mandel} in the early 1980's the interest of physicists has been attracted by this subject and caused a first wave of publications in the area of fractional Brownian motion and anomalous diffusion processes. But these works caused only minimal impact on the progress of traditional physics, because the results obtained could also be derived using classical methods \cite{samko}. 

This situation changed drastically by the progress made in the area of fractional wave equations during the last 10 years. Within this process, new questions in fundamental physics have been raised, which cannot be formulated adequately using traditional methods. Consequently a new research area has emerged, which allows for new insights und intriguing results using new methods and approaches.

The interest in fractional wave equations was reinforced in the year 2000 with a publication of Raspini \cite{raspini}. He demonstrated, that a 3-fold factorization of the Klein-Gordon equation leads to a fractional Dirac equation which contains fractional derivative operators of order $\alpha =2/3 $  and furthermore the resulting $\gamma$ - matrices obey an extended Clifford algebra.
To state this result more precisely: the extension of Dirac’s linearization procedure which determines the correct coupling of a SU(2) symmetric charge to a 3-fold factorization of the d'Alembert-operator leads to a fractional wave equation with an inherent SU(3) symmetry. This symmetry is formally deduced by the factorization procedure. In contrast to this formal derivation a standard Yang-Mills-theory is merely a recipe for coupling a phenomenologically deduced SU(3) symmetry.
Zavada \cite{zavada} has generalized Raspini's result: he demonstrated, that a n-fold factorization of the d'Alembert-operator automatically leads to fractional wave equations with an inherent SU(n) symmetry.

In 2002 Laskin \cite{laskin} on the basis of the Riesz definition \cite{riesz} of the fractional derivative presented a Schr\"odinger equation with fractional derivatives and gave a proof of Hermiticity and parity conservation of this equation.
In 2005 we \cite{he2005}  calculated algebraically the Casimir operators and multiplets of the fractional extension of the standard rotation group SO(n). This may be interpreted as a first approach
to investigate a fractional generalization of a standard Lie-algebra. The classification scheme derived was used for a successful description of the charmonium spectrum. The derived symmetry has been used to predict the exact masses of Y(4260) and X(4664), which later have been confirmed experimentally.
In the year 2006 Goldfain \cite{gold2006}  demonstrated, that the low level fractional dynamics \cite{tar2006} on a flat Minkowski metric most probably describes similar phenomena as a field theory in curved Riemann space time. In addition, he proposed a successful mechanism to quantize fractional free fields.
In 2006 Lim \cite{lim2006}  proposed a quantization description for free fractional fields of Klein-Gordon-type and investigated the temperature dependence of those fields.

In 2007 we \cite{her2007} applied the concept of local gauge invariance to fractional free fields  and derived the exact interaction form in first order of the coupling constant. The fractional analogue of the normal Zeeman-effect was calculated and as a first application a mass formula was presented, which gives the masses of the baryon spectrum with an accuracy better than 1$\%$.
It has been demonstrated, that the concept of local gauge invariance determines
the exact form of interaction, which in lowest order coincides
with the derived group chain for the fractional rotation group.
Furthermore we investigated the transformation properties of the fractional Schr\"odinger equation under rotations  with the result, that the particles described carry an additional intrinsic mixed rotational- and translational degree of freedom, which we named fractional spin. As a consequence the transformation properties of a fractional Schr\"odinger equation are closely related to a standard Pauli equation.
Since then  the investigation of the fractional rotation group within the framework of fractional group theory has lead to a vast amount of interestening results, e.g. a theoretical foundation of magic numbers in atomic nuclei \cite{he101} and metallic clusters \cite{he102}.

Like q-deformed Lie algebras the fractional rotation group extends the standard rotational symmetry and describes rotational and
vibrational degrees of freedom simultaneously.
There is a common aspect in the concepts of fractional calculus and q-deformed Lie algebras: they both extend the definition of the standard derivative operator.

Therefore we will try to establish a connection between the
q-derivative(\ref{qd}) 
and the fractional derivative definition.

The definition of the fractional order derivative is not unique,  several definitions 
e.g. the Riemann, Caputo, Weyl, Riesz, Feller,  Gr\"unwald  fractional 
derivative definition coexist \cite{caputo}-\cite{pod}.

But to be conformal with the requirements (\ref{q00}) and 
(\ref{qdx}), 
we may  apply the 
Caputo derivative $D^\alpha_x$:
\begin{equation}
\label{caputo}
D^{\alpha}_xf(x)=\cases{\frac{1}{\Gamma(1 -\alpha)}   
     \int_0^x  \!\! d\xi \, (x-\xi)^{-\alpha} \frac{\partial}{\partial \xi}f(\xi)\!\!\!\!\!\! & \qquad$0 \leq \alpha < 1$ \cr \\
\frac{1}{\Gamma(2 -\alpha)}  
     \int_0^x \!\!  d\xi \, (x-\xi)^{1-\alpha}  \frac{\partial^2}{\partial \xi^2}f(\xi)\!\!\!\!\!\!& \qquad$1 \leq \alpha < 2$ \cr }
\end{equation}
For a function set $f(x)= x^{n \alpha} $ we obtain:
\begin{equation}
\label{fx}
D^{\alpha}_x x^{n \alpha} = \cases{
{\Gamma(1+n \alpha) \over \Gamma(1+(n-1)\alpha)} x^{(n-1)\alpha} & $n>0$ \cr \\
0 &  $n=0$\cr}
\end{equation}
Let us  now interpret the fractional derivative parameter $\alpha$ as a deformation parameter in the sense of 
q-deformed Lie algebras. Setting $|n>= x^{n \alpha}$ we define:
\begin{equation}
\label{defa}
\left[ n \right]_\alpha |n>= \cases{
{\Gamma(1+n \alpha) \over \Gamma(1+(n-1)\alpha)} |n> &  $n>0$ \cr \\
0 &  $n=0$\cr}
\end{equation}
Indeed it follows
\begin{equation}
\label{q2limit}
\lim_{\alpha \rightarrow 1}\left[ n \right]_\alpha = n 
\end{equation}
The more or less abstract q-number is now interpreted within the mathematical context of fractional calculus
as the fractional derivative parameter $\alpha$ with a well understood meaning.

The definition (\ref{defa}) looks just like one more definition for a q-deformation. But there is a significant
difference which makes the definition based on fractional calculus unique. 

Standard q-numbers are defined more or less heuristically. There exists no physical or mathematical framework, which
determines their explicit structure. Consequently, many different definitions have been 
proposed in literature (see e.g. \cite{bon}).  

In contrast to this diversity, the q-deformation based on the definition of the
fractional derivative is uniquely determined, once a set of basis functions is given.

As an example we will derive the q-numbers for the fractional harmonic oscillator in the next section.
\section{The fractional q-deformed harmonic oscillator}
In the following we will present a derivation of a fractional q-number, which is independent from the specific choice of a fractional derivative definition.

The transition from classical mechanics to quantum mechanics may be
performed by canonical quantisation \cite{dirac}, \cite{messiah}. The classical canonically conjugated 
observables $x$ and $p$ are replaced by quantum mechanical observables
$\hat{x}$ and $\hat{p}$, which are introduced  as derivative operators on a Hilbert
space of square integrable wave functions $f$, see e.g. \cite{he}. The space coordinate representations 
of these operators are
\begin{eqnarray}
\label{classic1}
\hat{X} f(x) &= x f(x) \\
\label{classic2}
\hat{P} f(x) &= -i \hbar \partial_x f(x)
\end{eqnarray}
where $\hat{X}$ and $\hat{P}$ fulfill the commutation relation
\begin{equation}
\label{comm0}
\left[ \hat{X}, \hat{P} \right] = i \hbar 
\end{equation}
 In fractional calculus, the classical derivatives are 
extended to fractional derivative definitions. Using the 
fractional derivative $D^\alpha_{x} $, the space coordinate
representation of these operators is given by \cite{he}: 
\begin{eqnarray}
\label{f1}
  \hat{x} &=&  \left( \frac{\hbar}{m c} \right)^{(1-\alpha)} x^\alpha  \\
\label{f2}
 \hat{p} &=& -i \left( \frac{\hbar}{m c} \right)^{\alpha} m c \, D^\alpha_{x} 
\end{eqnarray}
The attached factors $(\hbar/m c)^{(1-\alpha)}$ and $(\hbar/m c)^\alpha m c$ 
respectively  ensure correct length and momentum units. 
The 
commutation relations $\left[ \hat{x}, \hat{p} \right]$ become $\alpha$ dependent \cite{he}.  
Of course, for  $\alpha=1$ these operators reduce to the standard $\hat{X}$ and $\hat{P}$.

With these operators, the classical Hamilton function for the harmonic oscillator
\begin{equation}
H_{\rm{class}} = {p^2 \over 2 m} + \frac{1}{2} m \omega^2 x^2
\end{equation}
is quantized. This yields the Hamiltonian $H^\alpha$
\begin{equation}
H^\alpha= {\hat{p}^2 \over 2 m} + \frac{1}{2} m \omega^2 \hat{x}^2
\end{equation}
and the corresponding stationary Schr\"odinger equation is explicitely given by:
\begin{equation}
H^\alpha \Psi = \bigl( - {1 \over 2 m} 
\left( \frac{\hbar}{m c} \right)^{2 \alpha} m^2 c^2 \, D^\alpha_{x}D^\alpha_{x}
+ \frac{1}{2}m \omega^2 \left( \frac{\hbar}{m c} \right)^{2(1-\alpha)} x^{2 \alpha}
\bigr) \Psi
 =  E \Psi
\end{equation}
It should be emphasized, that the Hermiticity of the proposed Hamiltonian depends on the specific choice of a fractional derivative definition. While for the Caputo- and Riemann definition of the fractional derivative the resulting Hamiltonian is not hermitean, it can be shown, that the use of the Feller- and Riesz definition of the fractional derivative assures Hermiticity of the Hamiltonian $H^\alpha$ \cite{laskin}.

Introducing the variable $\xi$ and the scaled energy $E'$:
\begin{eqnarray}
\xi^\alpha  &=& \sqrt{\frac{m \omega}{\hbar}} 
\left( \frac{\hbar}{m c} \right)^{1- \alpha} x^\alpha \\
E & = & \hbar \omega E'
\end{eqnarray}
we obtain the stationary Schr\"odinger equation for the fractional harmonic oscillator in the  canonical form
\begin{equation}
\label{fho}
H^\alpha \Psi_n(\xi) = {1 \over 2 }\bigl( - D^{2 \alpha}_{\xi}+ \xi^{2 \alpha}
\bigr)\Psi_n(\xi)
 =  E'(n,\alpha)  \Psi_n(\xi)
\end{equation}
Laskin \cite{laskin} has derived an approximate analytic solution within the framework of the
WKB-approximation, which has the advantage, that it is independent of the
choice of a specific definition of the  fractional derivative. We adopt his result:   
\begin{equation}
E'(n,\alpha)  =  
\bigl(
{1 \over 2} +n
\bigr)^\alpha
\pi^{\alpha/2}\left({\alpha \Gamma(\frac{1+\alpha}{2 \alpha}) \over \Gamma(\frac{1}{2 \alpha})}\right)^\alpha  
\quad n=0,1,2,...
\end{equation}

 In view of q-deformed Lie-algebras, we can use this analytic result to derive the corresponding q-number. With
 (\ref{eho}) the q-number is determined by the recursion relation:
\begin{equation}
\left[ n+1 \right]_\alpha = 2 E'(n,\alpha) - \left[ n \right]_\alpha 
\end{equation}
The obvious choice $\left[ 0 \right]_\alpha = 0$ for the initial condition leads to an oscillatory behaviour for 
$\left[ n \right]_\alpha$ for $\alpha < 1$. If we require a monotonically increasing behaviour of $\left[ n \right]_\alpha$ for
increasing $n$, an adequate choice for the 
initial condition is
\begin{equation}
\left[ 0 \right]_\alpha =
2^{1+\alpha}
\pi^{\alpha/2}\left({\alpha \Gamma(\frac{1+\alpha}{2 \alpha}) \over \Gamma(\frac{1}{2 \alpha})}\right)^\alpha  
\bigl(
 \zeta(-\alpha,\frac{1}{4})
-\zeta(-\alpha,\frac{3}{4})
\bigr)
\end{equation}
The explicit solution is then given by:
\begin{equation}
\label{hoq}
\left[ n \right]_\alpha  =  
2^{1+\alpha}
\pi^{\alpha/2}\left({\alpha \Gamma(\frac{1+\alpha}{2 \alpha}) \over \Gamma(\frac{1}{2 \alpha})}\right)^\alpha  
\bigl(
 \zeta(-\alpha,\frac{1}{4} + \frac{n}{2})
-\zeta(-\alpha,\frac{3}{4} + \frac{n}{2})
\bigr)
\end{equation}
where $\zeta(s,x)$ is the incomplete Riemann or Hurwitz zeta function, which is defined as:
\begin{equation}
 \zeta(s,x) = \sum_{k=0}^\infty (k+x)^{-s}
\end{equation}
and for $\alpha=m \in 0,1,2,...$ it is related to the Bernoulli polynomials $B_m$ via
\begin{equation}
 \zeta(-m,x) =- {1 \over (m+1)}B_{m+1}(x)
\end{equation}
\begin{figure}
\begin{center}
\includegraphics[type=eps, ext=.eps, read=.eps, width=80mm,height=59mm]{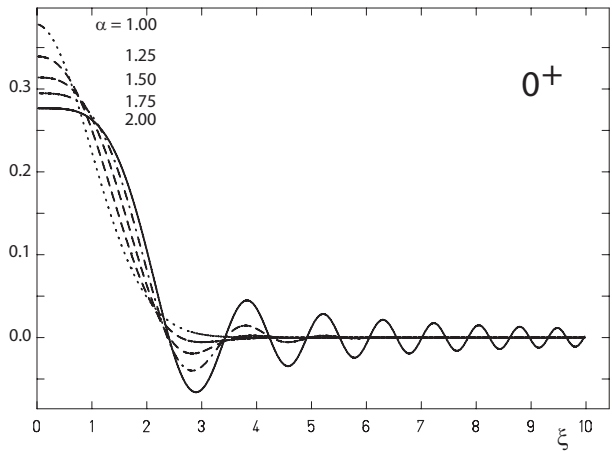}\\
\caption{\label{gs}
Plot of the  ground state wave function $\Psi_{0^+}(\xi)$ of the fractional harmonic oscillator (\ref{fho}), solved
with the Caputo fractional derivative (\ref{caputo}) for different
$\alpha$. 
} 
\end{center}
\end{figure}
Of course, the vacuum state $|0>$ is no more characterized by a vanishing expectation value of the
annihilation operator, but is defined by a zero expectation value of the number operator, which is
the inverse function of the fractional q-number (\ref{hoq}):
\begin{equation}
N |0> = \bigl(  \left[ n \right]_\alpha \bigr)^{-1}  |0> = n |0> = 0
\end{equation}

Following an idea of Goldfain \cite{gol08} we may interpret the fractional derivative as a simultaneous description of a particle and a corresponding gauge field. Interpreting the vacuum state as a  particles absent, but  gauge field present state,  the non-vanishing expectation value of the annihilation operator indicates the presence of the gauge field, while the number operator only counts for particles.

Setting  $\alpha=1$ leads to $\left[ n \right]_{\alpha=1} = n$ and consequently, the standard harmonic oscillator
energy spectrum $E'(n,\alpha=1) = (1/2+n)$ results. 
For $\alpha=2$ it follows
\begin{eqnarray}
E'(n,\alpha=2) &=& 4 \pi\left({\Gamma(\frac{3}{4}) \over \Gamma(\frac{1}{4})}\right)^2 \,(\frac{1}{2}+n)^2\\
               &=& {8 \pi^3 \over  \Gamma(\frac{1}{4})^4}\, (\frac{1}{4}+ n + n^2)\\
               &=& {2 \pi^3 \over  \Gamma(\frac{1}{4})^4} +
                  {8 \pi^3 \over  \Gamma(\frac{1}{4})^4}
                                                \bigl( n(n+1)\bigr)
\end{eqnarray}
which matches, besides a non vanishing zero point energy contribution, a  spectrum of rotational type
$E_{\rm{rot}}\equiv l(l+1), \quad l=0,1,2,..$. 

Unlike applications of ordinary q-numbers \cite{be2}, this 
result is not restricted to a finite number $n$, but is valid for all $n$. This is a significant enhancement and allows to apply this model for high energy excitations, too.

Furthermore, it should at least be mentioned, that the above few lines mark the realization of an old alchemist's dream: the transmutation of a given group to another (here from U(n) to SO(3)). This was  the ambitious aim of q-deformed Lie algebras \cite{gup} for decades, but was missed until now. 

\begin{figure}
\begin{center}
\includegraphics[type=eps, ext=.eps, read=.eps,width=130mm,height=49mm]{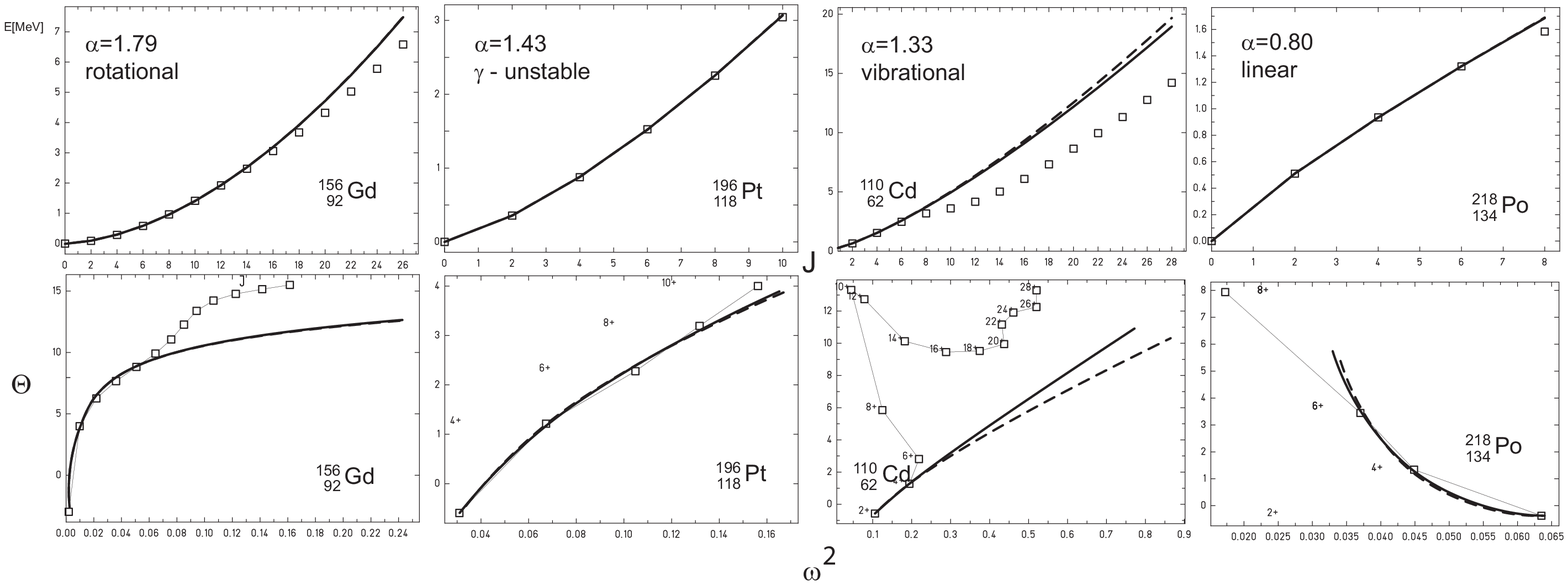}\\
\caption{
\label{rotgamvib}
The upper row shows energy levels of the ground state bands  for 
 $^{156}$Gd, $^{196}$Pt, $^{110}$Cd and $^{218}$Po, which are plotted for increasing $J$.  Squares
indicate the experimental values given in \cite{156GD}-\cite{218PO}, 
The full line indicates the optimum fit for the fractional q-deformed harmonic oscillator using (\ref{evib}) and dashed lines give
the best fit for the fractional q-deformed ${\rm{SU}}_\alpha(2)$ symmetric rotor according to (\ref{erot}). 
In the lower row the corresponding backbending plots are shown, which allow for the determination of 
the maximum angular momentum $J_{\rm{max}}$ for a valid fit  below the onset of alignment effects. 
} 
\end{center}
\end{figure}

In figure ({\ref{gs}}) we have plotted the ground state wave function of the 
fractional harmonic oscillator (\ref{fho}), obtained numerically for the Caputo fractional derivative (\ref{caputo}) for 
different $\alpha$. While for $\alpha \approx 1$ the wave function behaves like $\exp(-\xi^2/2)$, in the
vicinity of $\alpha \approx 2$ it looks more like a Bessel type function.

We obtained two idealized limits for the energy spectrum of the fractional harmonic oscillator. 
For $\alpha=1$ a vibration type spectrum is generated, while for $\alpha=2$ a rotational spectrum 
results. 
The fractional
derivative coefficient $\alpha$ acts like an order parameter and allows for a smooth transition between these  idealized
limits. 
Therefore the properties of the fractional harmonic oscillator seem to be well suited to describe e.g. the ground state
band spectra of even-even nuclei.
 We define
\begin{equation}
\label{evib}
E^{\rm{vib}}_J(\alpha,m_0^{\rm{vib}},a_0^{\rm{vib}}) = m_0^{\rm{vib}} + a_0^{\rm{vib}}(\left[J\right]_\alpha+ \left[J+1\right]_\alpha)/2 \qquad J=0,2,4,...
\end{equation}
where $m_0^{\rm{vib}}$ mainly acts as a counter term for the zero point energy and $a_0^{\rm{vib}}$ is a measure
for the level spacing and use (\ref{evib}) for a fit of the experimental ground state band spectra
of
 $^{156}$Gd, $^{196}$Pt, 
$^{110}$Cd and $^{218}$Po  which represent typical rotational-, $\gamma$-unstable-, vibrational and linear 
type spectra \cite{156GD}-\cite{218PO}. 

In table \ref{tabvib} the optimum parameter sets for a fit with the experimental data are listed. In figure \ref{rotgamvib} the results are plotted.  Below the onset of microscopic alignment effects
all spectra are described with an accuracy better than $2\%$. Therefore the fractional q-deformed harmonic
oscillator indeed describes the full variety of ground state bands of even-even nuclei with remarkable
accuracy.

The fractional derivative coefficient $\alpha$ acts like an order parameter and
allows a smooth transition between these idealised cases. 

Of course, a smooth
transition between rotational and e.g. vibrational spectra may be achieved by geometric
collective models, too. A typical example is the Gneuss-Greiner model \cite{gg} with a more
sophisticated potential. Critical phase transitions from vibrational to rotational states 
have been studied for decades using e.g. coherent states formalism or 
within the framework of the IBA-model \cite{gg2}-\cite{gg5}.  
But in general, within these models, results may be obtained only numerically 
with extensive effort. 

The proposed fractional q-deformed harmonic oscillator is a fully analytic model, which may be applied with minimum effort not only to the limiting idealized cases of the pure rotor and pure vibrator respectively, but covers all intermediate configurations as well.

\section{The fractional q-deformed symmetric rotor}
In the previous section we have derived the q-number associated with the  fractional harmonic oscillator. 
Interpreting equation (\ref{hoq}) as a formal definition, the Casimir-operator $C(\rm{SU}_\alpha(2))$
\begin{equation}
C(\rm{SU}_\alpha(2)) = \left[ J \right]_\alpha \left[J+1\right]_\alpha \quad J=0,1,2,...
\end{equation}
of the group $\rm{SU}_\alpha(2)$ is determined. This group is generated by the operators $J_+$, $J_0$ and
$J_-$, satisfying the commutation relations:
\begin{eqnarray}
\left[ J_0, J_\pm \right] &=& \pm J_\pm    \\
\left[ J_+, J_- \right] &=& \left[2 J_0\right]_\alpha    
\end{eqnarray}
Consequently we are able to define the fractional q-deformed symmetric rotor as
\begin{equation}
\label{erot}
E^{\rm{rot}}_J(\alpha,m_0^{\rm{rot}},a_0^{\rm{rot}}) = m_0^{\rm{rot}} + a_0^{\rm{rot}}\left[J\right]_\alpha  \left[J+1\right]_\alpha \quad J=0,2,4,...
\end{equation}
where $m_0^{\rm{rot}}$ mainly acts as a counter term for the zero point energy and $a_0^{\rm{rot}}$ is a measure
for the level spacing.
\begin{table}
\caption{Listed are the optimum parameter sets ($\alpha$, $a_0^{\rm{vib}}$, $m_0^{\rm{vib}}$ according (\ref{evib})) 
for the fractional
harmonic oscillator for different nuclids. The maximum
angular momentum $J_{max}$ for a valid fit below the onset of alignment effects is given as well as the root mean square 
error $\Delta$E between experimental (\cite{156GD}-\cite{218PO})
 and fitted energies in $\%$. 
 }
\label{tabvib}       
\begin{tabular}{llrrrrr}
\hline\noalign{\smallskip}
nuclid & $\alpha$ & $a_0^{\rm{vib}}[keV]$ & $m_0^{\rm{vib}}[keV]$ & $J_{max}$ & $\Delta$E[$\%$]  \\
\noalign{\smallskip}\hline\noalign{\smallskip}
$^{156}_{\,\,\,92}$Gd$_{64} $  &1.795 &  15.736 &  -14.136 & 14 & 1.48 \\ 
$^{196}_{118}$Pt$_{78}$  &1.436 &  91.556 &  -39.832 & 10 & 0.16\\ 
$^{110}_{\,\,\,62}$Cd$_{48} $  &1.331 & 197.119 &  -87.416 &  6 & 0 \\ 
$^{218}_{134}$Po$_{84}$  &0.801 & 357.493 & -193.868 &  8 & 0.06 \\ 
\end{tabular}
\end{table}

For $\alpha=1$, $E^{\rm{rot}}$ reduces to $E^{\rm{rot}} = m_0^{\rm{rot}} + a_0^{\rm{rot}}J(J+1)$, which is the 
spectrum of a symmetric rigid rotor. For $\alpha=1/2$ we obtain:
\begin{equation}
\lim_{J \rightarrow \infty} \, (E^{\rm{rot}}_{J+2} -  E^{\rm{rot}}_{J}) =  a_0^{\rm{rot}}\pi/2 = {\rm{const}}
\end{equation}
which is the spectrum of a harmonic oscillator. 
We define the ratios
\begin{eqnarray}
R^{\rm{vib}}_{J,\alpha} &=& {(E^{\rm{vib}}_{J}-E^{\rm{vib}}_{0}) \over (E^{\rm{vib}}_{2}-E^{\rm{vib}}_{0})}  \\
R^{\rm{rot}}_{J,\alpha} &=& {(E^{\rm{rot}}_{J}-E^{\rm{rot}}_{0}) \over (E^{\rm{rot}}_{2}-E^{\rm{rot}}_{0})} 
\end{eqnarray}
which only depend on $J$ and $\alpha$. 
A Taylor series expansion at $J=2$ and $\alpha=1$ leads to:  
\begin{eqnarray}
R^{\rm{vib}}_{J,\alpha} &=& 1 + (J-2) (0.10 - 0.04 (J-2)) \nonumber \\ & & 
+ (\alpha-1)(J-2)(0.101   - 0.020  (J-2))  \\
R^{\rm{rot}}_{J,\alpha/2} &=& 1 + (J-2) (0.11  - 0.04 (J-2)) \nonumber \\ & & 
 + (\alpha-1)(J-2)(0.095  - 0.016  (J-2)) 
\end{eqnarray}
A comparison of these series leads to the remarkable result
\begin{equation}
\label{xchange}
R^{\rm{vib}}_{J,\alpha} \simeq R^{\rm{rot}}_{J,\alpha/2}  \qquad + o(J^3,\alpha^2)
\end{equation}
Therefore the fractional q-deformed harmonic oscillator (\ref{evib}) and the fractional
q-deformed symmetric rotor (\ref{erot}) generate similar spectra. As a consequence, a fit of the experimental
ground state band spectra of even-even nuclei with (\ref{evib}) and (\ref{erot}) respectively leads to similar results,
as demonstrated in figure \ref{rotgamvib}. There is no difference between rotations
and vibrations any more, the corresponding spectra are mutually connected via relation (\ref{xchange}).

\begin{figure}
\begin{center}
\includegraphics[type=eps, ext=.eps, read=.eps,width=80mm,height=58mm]{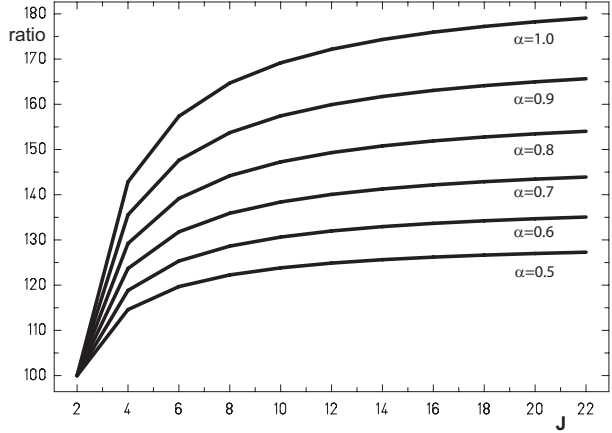}\\
\caption{
\label{fbe2}
B(E2)-values for the fractional q-deformed ${\rm{SU}}_\alpha(2)$ symmetric rotor according to (\ref{be2}), 
normalized with $100 / B_\alpha(E2;2^+ \rightarrow 0^+)$.
} 
\end{center}
\end{figure}

Finally we may consider the behaviour of $B(E2)$-values for the fractional q-deformed symmetric rotor. Using the formal equivalence with q-deformation,  these values are given by \cite{be2}:
\begin{equation}
\label{be2}
B_\alpha(E2;J+2 \rightarrow J) = 
{5 \over 16 \pi} Q_0^2
{
\left[ 3 \right]_\alpha
\left[ 4 \right]_\alpha
\left[ J+1 \right]_\alpha^2
\left[ J+2 \right]_\alpha^2
\over
\left[ 2 \right]_\alpha
\left[ 2 J+2 \right]_\alpha
\left[ 2 J+3 \right]_\alpha
\left[ 2 J+4 \right]_\alpha
\left[ 2 J+5 \right]_\alpha
}
\end{equation}
In figure \ref{fbe2} these values are plotted, normalized with $100 / B_\alpha(E2;2^+ \rightarrow 0^+)$. Obviously
there is a saturation effect for increasing $J$.

\section{Half integer representations of the fractional rotation group}
Up to now we have investigated the ground state 
excitation spectra of even-even nuclei. The next step of 
complication arises if the proton or neutron number is odd.
In classical models it is expected, that excitations of both collective and single-particle
type will be possible and in general will be coupled.
Approximately, the even-even nucleus is treated as a collective core, whose
internal structure is not affected by one more particle moving on its surface.
In the strong-coupling model, the corresponding strong coupling Hamiltonian $H_{sc}$ is 
decomposed into a collective, single-particle and  interaction term \cite{eg}:
\begin{equation}
H_{sc} = H_{coll}^0 + H_{sp}^0 + H_{ii}
\end{equation}
For $K=1/2$ ground state bands the energy level spectrum  is known analytically \cite{mar}:
\begin{equation}
E_{K=1/2}(I) = m_0 + c_0 I(I+1) + a_0 (-1)^{I+1/2}(I+1/2) \quad I=1/2,3/2,5/2,...
\end{equation}
where $a_0$ is called the decoupling parameter and $m_0$ and $c_0$ in units (keV) are parameters to be fitted
with experimental data. 

In view of a q-deformed Lie-algebra of the standard rotation group $SO(3)$ this result may be interpreted as an expansion in terms of of Casimir-operators:
\begin{eqnarray}
E_{K=1/2}(I) &=& m_0 + c_0 C(SO(3)_q) + a_0 (-1)^{I+1/2}C(SO(2)_q) \\
 &=& m_0 + c_0 [I]_q[I+1]_q + a_0 (-1)^{I+1/2}[I+1/2]_q
\end{eqnarray}
Consequently the investigation of the ground state 
band spectra of even-odd and odd-even nuclei respectively is 
a test of whether or not  half-integer representations of the fractional q-deformed
rotation group are realized in nature.

Furthermore, in view of fractional calculus, we may use the
explicit form (\ref{hoq}) of the q-number associated with the
fractional harmonic oscillator:
\begin{eqnarray}
E_{K=1/2}(I) &=& m_0 + c_0 C(SO(3)_\alpha) + a_0 (-1)^{I+1/2}C(SO(2)_\alpha) \\
\label{ek}
 &=& m_0 + c_0 [I]_\alpha[I+1]_\alpha + a_0 (-1)^{I+1/2}[I+1/2]_\alpha
\end{eqnarray}
\begin{figure}
\begin{center}
\includegraphics[type=eps, ext=.eps, read=.eps,width=80mm,height=58mm]{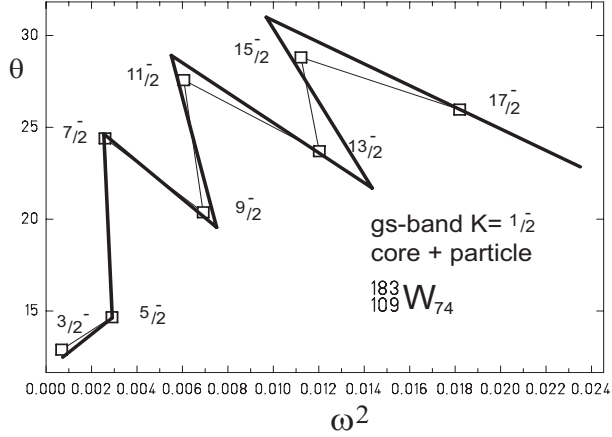}\\
\caption{
\label{fww}
backbending plot for $^{183}$W$_{74}$ for the $K=1/2^-$ ground state band. Thick line is the fit result according (\ref{ek}). Squares indicate the experimental values. 
}
\end{center}
\end{figure}
In figure \ref{fww} we show a fit of the experimental values for the $K=1/2^{-}$ band of 
the nucleus  $^{183}$W$_{74}$ with (\ref{ek}). With
$\alpha=1.02$, $m_0= -6.077$, $c_0=11.76$ and $a_0 = 2.09$ the rms-error is less than $1 \%$. This is a first indication,
that half-integer representations of the fractional
rotation group may be indeed realized in nature.

\section{Conclusion}
We have shown, that fractional calculus and q-deformed Lie algebras are closely related. 
Both concepts expand the scope of standard Lie algebras to describe generalized symmetries. 
While a standard q-deformation is merely introduced heuristically and in general no underlying mathematical motivation
is given, the q-deformation based on the definition of the
fractional derivative is uniquely determined, once a set of basis functions is given. For the
fractional harmonic oscillator, we  derived the corresponding q-number and  demonstrated, that
the resulting energy spectrum is an appropriate tool  to describe e.g. the ground state spectra of even-even nuclei. 

In addition, the equivalence of rotational and vibrational spectra 
for fractional q-deformed Lie algebras has been demonstrated and we have calculated the $B_\alpha(E2)$ values
for the fractional q-deformed symmetric rotor.

The results derived encourage further studies in this field.
The application of fractional calculus to phenomena, which until now have been described using
q-deformed Lie algebras will lead to a broader
understanding of the underlying generalized symmetries.
Furthermore fractional q-deformed Lie algebras indeed allow a smooth transition between different standard Lie algebras and therefore open up a new field of research.

\section{Acknowledgments}
We thank A. Friedrich for  useful discussions.

\begin{thebibliography}{00}
%
%
\bibitem{bon} Bonatsos D and Daskaloyannis C,
{\it Quantum groups and their applications in nuclear physics},
 Prog. Part. Nucl. Phys. {\bf 43} (1999) 537-618
\bibitem{ol} Oldham K B and Spanier J (2006), 
{\it The Fractional Calculus},
Dover Publications, Mineola, New York.
\bibitem{hil08}  
Hilfer R,
{\it Threefold introduction to fractional drivatives}, 
in: Klages R. ,  Radons G. and  Sokolov I. (editors). 
Anomalous Transport, 
Foundations and Application,
Wiley-VCH, Weinheim, Germany, (2008) pp. 17 
\bibitem{mai10}
 Mainardi F, (2010)
{\it  Fractional Calculus and Waves in Linear Viscoelasticity: An Introduction to Mathematical Models}, World Scientific, Singapore.
\bibitem{he08}  Herrmann R (2008) {\it Fraktionale Infinitesimalrechnung - Eine Einf\"uhrung f\"ur Physiker}, BoD, Norderstedt, Germany
\bibitem{baleanu10} 
Baleanu D and Trujillo J (2010). 
{\it A new method of finding the fractional Euler-Lagrange and Hamilton equations within Caputo fractional derivatives},
Comm. Nonlin. Sci. Num. Sim. {\bf 15}(5) (2010) 1111
\bibitem{ff}  Miller K  and  Ross B (1993)  {\it An Introduction to Fractional Calculus and Fractional Differential
Equations}, Wiley, New York.
\bibitem{abel} Abel N H, 
{\it Solution de quelques problemes a l’aide d’integrales definies}, Oevres Completes, Vol. 1, Grondahl, Christiana, Norwegen, pp.16-18
\bibitem{mandel}  Mandelbrot B, {\it Fractal geometry of nature}, Birkhaeuser (1991)
\bibitem{samko}  Samko S, Lebre A and Dos Santos A F (Eds.), {\it Factorization, Singular Operators and Related Problems}, Proceedings of the Conference in Honour of Professor Georgii Litvinchuk Springer Berlin (2003), pp. 152
\bibitem{raspini}Raspini A, 
{\it Dirac equation with fractional derivatives of order 2/3}, 
Fizika B {\bf9} (2000) 49
\bibitem{zavada}Zavada P J, 
{\it Relativistic wave equations with fractional derivatives and pseudo-differential operators}, 
J. Appl. Math. {\bf2}(4) (2002) 164
\bibitem{laskin}Laskin N, 
{\it Fractional Schr\"odinger equation}, 
Phys. Rev. E {\bf 66} (2002) 056108
\bibitem{he2005}Herrmann R, 
{\it Continuous differential operators and a new interpretation of the charmonium spectrum}, arxiv:nucl-th/0508033 and 
  Herrmann R, 
{\it Properties of a fractional derivative Schroedinger type wave equation and a new interpretation of the charmonium spectrum}, arxiv:math-ph/0510099

\bibitem{gold2006}Goldfain E, 
{\it Complexity in quantum-field theory and physics beyond the standard model}, 
Chaos, Solitons and Fractals 28, (2006) 913
\bibitem{tar2006}  Tarasov V E, Zaslavsky G M, 
{\it Dynamics with low-level fractionality}, 
Physica A \textbf{368} (2006) 399-415
\bibitem{lim2006}  Lim S C, 
{\it Fractional derivative quantum fields at finite temperatures}, Physica A, 363 (2006) 269 
\bibitem{her2007}Herrmann R, 
{\it Gauge invariance in fractional field theories}, 
Phys.Lett. A324 (2008) 5515
\bibitem{lim09} 
Lim S C,
{\it Repulsive Casimir force from fractional Neumann boundary conditions},
Phys. Lett. {\bf B679}, (2009) 130
\bibitem{he101}  
Herrmann  R,
{\it Higher-dimensional mixed fractional rotation groups as a basis for dynamic symmetries generating the spectrum of the deformed Nilsson oscillator},
Physica  A{\bf 389}(4) (2010)  693
\bibitem{he102}  
Herrmann R,   
{\it Fractional phase transition in medium size metal clusters},
Physica  A{\bf 389}(16) (2010) 3307
\bibitem{caputo} 
Caputo M,  
{\it Linear model of dissipation whose Q is almost frequency independent Part II},
Geophys. J. R. Astr. Soc.  \textbf{13}, (1967) 529.
\bibitem{weyl} 
Weyl H,  
{\it Bemerkungen zum Begriff des Differentialquotienten gebrochener Ordnung},
Vierteljahresschrift der Naturforschenden
Gesellschaft in Z\"urich  \textbf{62}, (1917) 296.
\bibitem{riesz} 
Riesz M,  
{\it L'integrale de Riemann-Liouville et le probl$\acute{\textrm{e}}$me de Cauchy},
Acta Math.   \textbf{81}, (1949) 1.
\bibitem{grun} 
Gr\"unwald A K, 
{\it \"Uber begrenzte Derivationen und deren Anwendung},
Z. angew. Math. und Physik   \textbf{12}, (1867) 441.
\bibitem{feller} Feller W, Comm. Sem. Mathem. Universite de Lund, (1952) 73-81.
\bibitem{kir} 
Kiryakova V S, 1994
{\it Generalized fractional calculus and applications},
 Longman (Pitman  Res. Notes in Math. Ser. {\bf 301}),
Harlow; co-publ.: J. Wiley and Sons, New York.
\bibitem{pod} Podlubny I, 1999 {\it Fractional Differential equations},
Academic Press, New York.
\bibitem{wu}
 Wu G C  and He J H, 
{\it Fractional Calculus of Variations in fractal spacetime}, 
Nonlinear Science Letters A, \textbf{1}(3)(2010) 281
\bibitem{dirac}  Dirac P A M,  1930  {\it The Principles of Quantum Mechanics} The Clarendon Press, Oxford.
\bibitem{messiah}  Messiah A, 1968  {\it Quantum Mechanics} John Wiley $\&$ Sons, North-Holland Pub. Co , New York. 
\bibitem{gol08}  
Goldfain E,
{\it
 Fractional dynamics and the TeV regime of field theory} Communications in non linear science and numerical simulation, 
{\bf 13} (2008) 666.
\bibitem{he}  Herrmann R, 
{\it The fractional symmetric rigid rotor},
J. Phys. G: Nucl. Part. Phys.  {\bf 34} (2007)  607
\bibitem{gup} 
Gupta  R~K, Cseh J, Ludu  A, Greiner W and Scheid W, 
{\it Dynamical symmetry breaking in SU$(2)$ model and the quantum group SU$(2)_q$}
J. Phys. G   \textbf{18} (1992)  L73.
\bibitem{156GD}
Reich C W,  Nuclear Data Sheets \textbf{99},  (2003) 753.
\bibitem{196PT}
Chunmei Z, Gongqing W and Zhenlan T,  Nuclear Data Sheets \textbf{83}, (1998) 145.
\bibitem{110CD}
Frenne D  and  Jacobs E,  Nuclear Data Sheets \textbf{89},  (2000) 481.
\bibitem{218PO}
Singh B,  Nuclear Data Sheets \textbf{107}, (2006) 1027.
\bibitem{be2}
Raychev P P, Roussev R P and Smirnov Yu F J, 
{\it The quantum algebra $SU_q(2)$ and rotational spectra of deformed nuclei },
Phys. G. Nucl. Part. Phys. \textbf{16},  (1990) L137.
\bibitem{gg}
Gneuss G  and Greiner W, 
{ \it Collective potential energy surfaces and nuclear structure}, 
Nucl. Phys. A\textbf{171}, (1971) 449.
\bibitem{gg2}
Casten R F, Zamfir N V and Brenner D S,  
{\it Universal anharmonic vibrator description of nuclei and critical nuclear phase transitions},
Phys. Rev. Lett. \textbf{71}, (1993) 227.
\bibitem{gg3}
Feng D H, Gilmore R and Deans S R,  
{\it Phase transitions and the geometric properties of the interacting boson model},
 Phys. Rev. C \textbf{23}, (1981) 1254.
\bibitem{gg4}
Haapakoski P, Honkaranta P and Lipas P O,  
{\it Projection model for ground bands of even-even nuclei},
Phys. Lett. B \textbf{31}, (1970) 493.
\bibitem{gg5}
Raduta A A, Gheorghe A C and Faessler A J, 
{\it Remarks on the shape transition from spherical to deformed gamma unstable nuclei},
Phys. G \textbf{31}, (2005) 337.
\bibitem{eg}
Eisenberg J M  and Greiner W, 1987 {\it Nuclear Models} North Holland,Amsterdam.
\bibitem{mar} Greiner W and Maruhn J, 1996 {\it Nuclear Models} Springer, Berlin 
\end{thebibliography}
%

\end{document}